\documentclass[conference]{IEEEtran}
\usepackage{cite}
\usepackage{amsmath,amssymb,amsfonts}
\usepackage{algorithmic}
\usepackage{graphicx}
\usepackage{textcomp}
\usepackage{xcolor}
\usepackage[ruled,vlined]{algorithm2e}
\def\BibTeX{{\rm B\kern-.05em{\sc i\kern-.025em b}\kern-.08em
    T\kern-.1667em\lower.7ex\hbox{E}\kern-.125emX}}
\begin{document}

\title{A Quantum Annealing Approach to Reduce Covid-19 Spread on College Campuses}

\author{\IEEEauthorblockN{James Sud}
\IEEEauthorblockA{\textit{Quantum Computing @ Berkeley} \\
\textit{University of California, Berkeley}\\
Berkeley, CA, USA \\
jamessud@berkeley.edu}
\and
\IEEEauthorblockN{Victor Li}
\IEEEauthorblockA{\textit{Quantum Computing @ Berkeley} \\
\textit{University of California, Berkeley}\\
Berkeley, CA, USA \\
lytz@berkeley.edu}
}

\maketitle

\begin{abstract}
Disruptions of university campuses caused by COVID-19 have motivated strategies to prevent the spread of infectious diseases while maintaining some level of in person learning. In response, the proposed approach recursively applied a quantum annealing algorithm for Max-Cut optimization on D-Wave Systems, which grouped students into cohorts such that the number of possible infection events via shared classrooms was minimized. To test this approach, available coursework data was used to generate highly clustered course enrollment networks  representing students and the classes they share. The algorithm was then recursively called on these networks to group students, and a disease model was applied to forecast disease spread. Simulation results showed that under some assumptions on disease statistics and methods of spread, the quantum grouping method reduced both the total and peak percentage of infected students when compared against random groupings of students. Scaling to larger networks, it is possible that this quantum annealer-assisted grouping approach may provide practical advantage over classical approaches. This paper, however, is strictly a proof-of-concept demonstration of the approach and is not intended to argue for a quantum speedup.
\end{abstract}

\section{Introduction}
Quantum annealing (QA) has been one of the flagship procedures for performing meaningful calculations on Noisy Intermediate Scale Quantum (NISQ) Devices. QA, introduced in \cite{kadowaki98, apolloni89, finnila94}, aims to find the ground state of a target quantum Hamiltonian from a more trivial starting Hamiltonian by smoothly interpolating between the two. A regime of interest for QA is Adiabatic Quantum Computation (AQC) \cite{farhi00}, which corresponds to the limit of slow transition between the initial and target Hamiltonian. In AQC, qubits remains in the instantaneous ground state of the evolving Hamiltonian, thereby ending in the ground state of the final Hamiltonian. Recently, quantum annealers from D-Wave systems have become commercially available \cite{johnson11}.

Any problem that can be formulated as a specific form of a binary objective function has the potential to be run on quantum annealers, and for many problems there exists efficient methods to convert to this form \cite{date19}. Previous research has been able to formulate and solve important problem instances from a wide array of scientific fields, such as combinatorial optimization \cite{chapuis18,santra14,farhi12,mcandrew20,martonak04}, physics \cite{harris18,brooke99,santoro02}, biology \cite{perdomo12,mulligan20}, economics \cite{orus19,rosenberg16,ding19}, and operations research \cite{neukart17}. Problem instances that are natively cast as binary objective functions, however, are a natural choice for the application of QA. Previous research into such problems has been conducted,  benchmarking the performance of QA against classical heuristics and algorithms \cite{trevisan12,goemans95,dunning18, sahni76}. This research has shown proven quantum speedups for a specific graph optimization problem \cite{somma12}, as well as evidence of a scaling advantage of QA over classical simulated annealing for specific classes of more general problems \cite{albash18, denchev16}. However, the investigation of quantum speedups over more general problem instance sets has led to more inconclusive or instance-dependent results \cite{ronnow14, hamerly19}. For the purposes of this paper, a comparison to classical methods is not of great importance, but rather the paper focuses on the quantum annealing method and simply compares to random groupings in order to demonstrate that the annealers are not simply returning random results for problems of the size analyzed.

While there has been recent research investigating applications of graph theory and classical graph optimization to reduce disease spread \cite{crnkovic21, alguliyev21, bhapkar20}, this paper specifically applies  the Maximum-Cut graph problem (Max-Cut) to epidemiology. Max-Cut is a combinatorial optimization problem that serves to find a partition of a given graph such that the number of edges crossing the partition is maximized. Max-Cut has been proven to be NP-Complete \cite{karp72}, and has been widely studied in both classical and quantum optimization literature \cite{trevisan12,goemans95,dunning18, farhi14}. The problem has a classical objective function which simply counts the number of edges that cross the chosen cut:
\begin{equation}
f(x) = \sum_{(i,j) \in E} x_i \oplus x_j \label{maxcut_classical}
\end{equation}
Where $E$ represents the edge set of the graph and $x_i$'s are binary variables representing which of the two groups a given node has been assigned to. This cost function can easily be cast into a quantum Ising Hamiltonian, using the relation $x = \frac{1}{2}(I-Z)$
\begin{equation}
\frac{1}{2} (I -\sum_{(i,j) \in E} Z_iZ_j) \label{maxcut_quantum}
\end{equation}
Where $Z_i Z_j$ refers to a Pauli matrix $Z$ acting on qubits $i$ and $j$.

Conveniently, quantum annealers from D-Wave Systems natively implement Quadratic Unconstrained Binary Optimization (QUBO) problems, to which Ising Hamiltonians can be trivially cast. Thus, Max-Cut defined on an arbitrary graph can be efficiently run on a D-Wave quantum annealer to obtain sample solutions, after mapping the nodes of the graph to qubits on the quantum annealer \cite{dwavesolver}.

This paper, motivated by the recent COVID-19 pandemic, investigates the recursive application of QA-assisted Max-Cut to group students into physically separated cohorts, with respect to networks specified by the physical interactions of students on a college campus. The efficacy of this method is tested via numerical simulations of disease outbreak. We hypothesize that this method can reduce the spread of an infectious disease compared to random groupings of students, based on the intuition that “cutting” the most edges from a network reduces the total number of physical interactions between students, thus reducing the opportunities for the spread of disease.

This paper is organized as follows: in Sec \ref{networks}, we describe our method for creating  course enrollment networks (CENs) and student interaction networks (SINs), which encapsulate physical interactions on university campuses. Sec \ref{annealing} fleshes out details on our recursive QA-assisted Max-Cut algorithm. Sec \ref{epidemics} describes our methods for simulating and testing disease spread. In Sec \ref{results}, we numerically simulate our approach and compare disease spread to the case of random groupings. Discussions and further directions are presented in Sec \ref{discussion}.

\section{Methods}\label{methods}

\subsection{Generating Networks}\label{networks}
Social networks, commonly used in psychology, sociology, and other fields, are a natural tool for describing a university population, as they can encapsulate a given student body and possible interactions within it. In the networks we generated, nodes represent students and bidirectional edges represent possible physical interactions (and thereby possible disease spread) between two nodes/students. 

\begin{figure}[htb]
\centerline{\includegraphics[width=5.5cm]{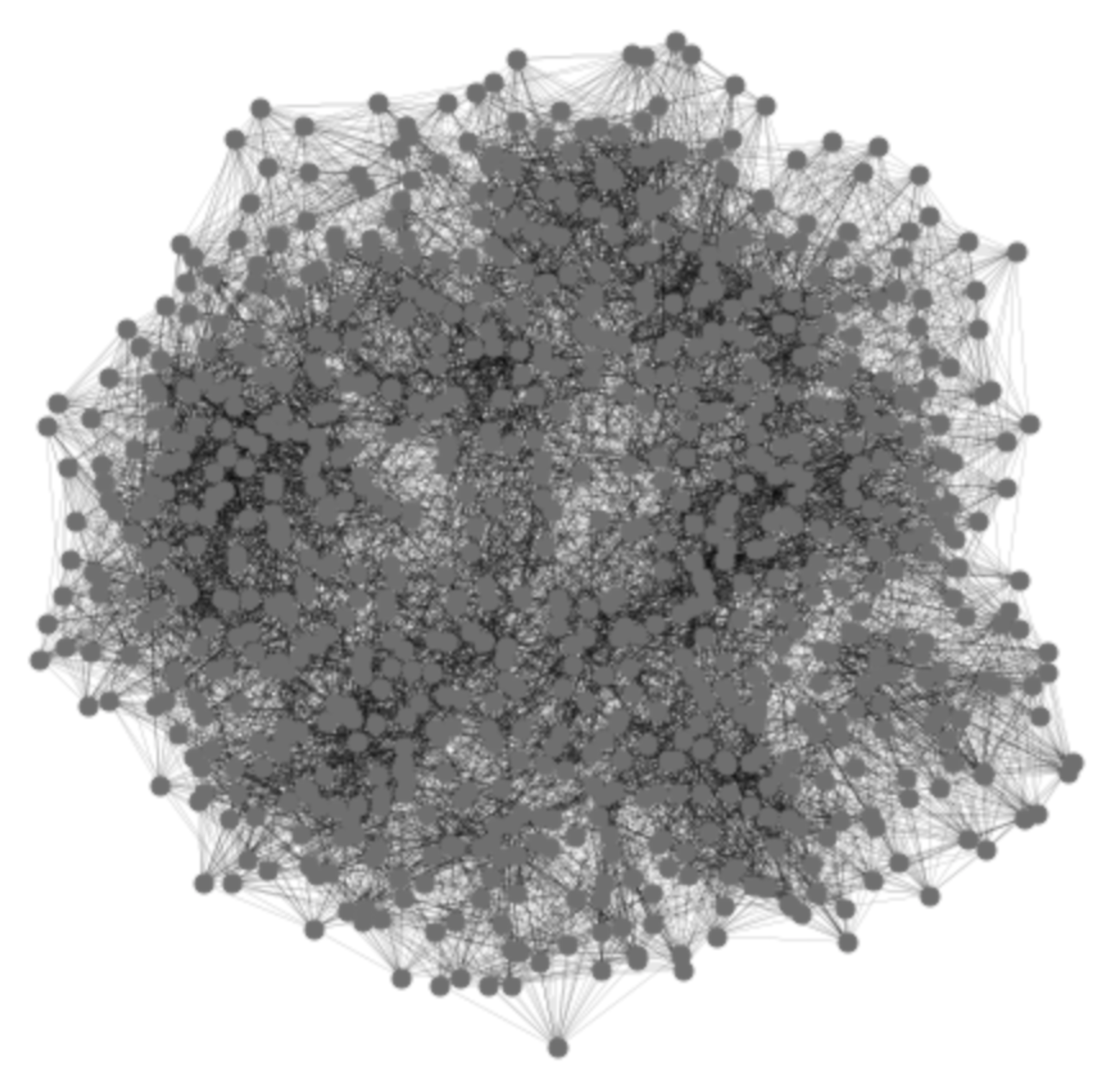}}
\caption{A sample Course Enrollment Network (CEN) based off data collected from Cornell University's Ithica Campus.}
\label{CEN}
\end{figure}

We first consider the case where two students share an edge only if they share a physical class in the same room at the same time, and refer to these networks as Course Enrollment Networks (CENs). To construct these CENs, we consulted published statistics of a social network based on the 2020 course enrollment statistics of Cornell University's Ithaca campus given in \cite{weeden20}. This reference specifies key network parameters listed in Table \ref{tab1}, and we found that, after toying with parameters, a Watts-Strogatz "small-world" graph \cite{watts98} generated using the NetworkX package \cite{networkx} produces a network with closest fit to these parameters. A sample CEN generated using this procedure can be found in Fig.~\ref{CEN}.

\begin{figure}[htb]
\centerline{\includegraphics[width=5.5cm]{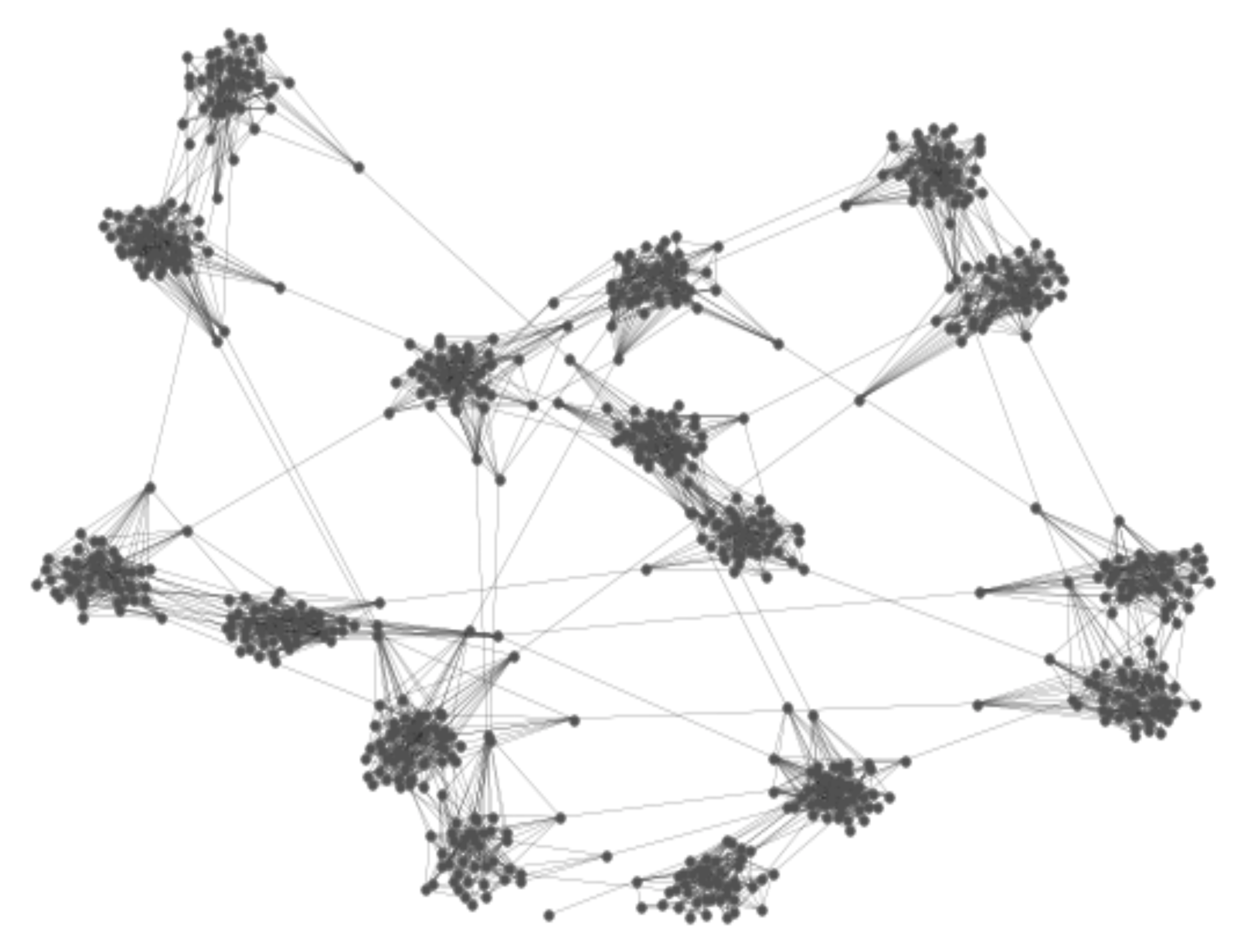}}
\caption{A sample Student Interaction Network (SIN) based on a returned assignment of 800 student to 16 cohorts (8 dormitories with 2 rooms each), with added floor, dorm, and campus interaction edges after partitioning.}
\label{SIN}
\end{figure}

In addition to CENs, we analyzed enhanced networks that more fully encapsulate the actual physical environment of a campus. To accomplish this, we added inter-cohort and intra-cohort interactions to our network with specified probabilities. We label these enhanced networks Student Interaction Networks (SINs). For our analysis of SINs, we considered a slightly smaller campus, modeled after Harvey Mudd College in Claremont, California. This college was chosen because approximately 99\% of the students live on campus \cite{harveymudd}, so we can attempt to more closely capture all possible interactions between students without considering off-campus influences. For the SINs, we assume that each cohort assigned by our algorithm lives in a separate designated dormitory hall or floor. Harvey Mudd College houses 800 students into 8 dorms, and each dorm has two floors on average \cite{harveymudd}. Thus, in order to create SINs with inter-cohort and intra-cohort interactions, we first generate a CEN of an 800 student campus, then divide the students into 16 groups (8 dorms and 2 floors), and add random edges between floors, dorms and dormitories with tunable probabilities. We label these edges floor, dorm, and campus interaction edges. We then run a disease model on the imperfectly separated cohorts. A sample SIN generated using this procedure can be found in Fig.~\ref{SIN}.

\begin{table}[htb]
\caption{Network parameters for CENs}
\begin{center}
\begin{tabular}{|c|c|c|}
\hline
\textbf{}&\textbf{\textit{Cornell}}& \textbf{\textit{NetworkX}} \\
\cline{2-3} 
\hline
\textbf{Number of nodes}&\textbf{\textit{3800}}& \textbf{\textit{3800}} \\
\hline
\textbf{Network density}&\textbf{\textit{0.040}}& \textbf{\textit{0.040}} \\
\hline
\textbf{Clustering coefficient}&\textbf{\textit{0.480}}& \textbf{\textit{0.465}} \\
\hline
\textbf{Mean geodesic distance}&\textbf{\textit{2.233}}& \textbf{\textit{3.110}} \\
\hline
\end{tabular}
\label{tab1}
\end{center}
\end{table}

\subsection{Network Partitioning}\label{annealing}
Once a CEN or SIN is defined, we can run a partitioning algorithm in order to divide students into cohorts. We outline Algorithm \ref{alg1}, which perform these partitions using a quantum annealer for the black-box subroutine \emph{performMaxcut}, and Algorithm \ref{alg2}, which randomly assigns each student to a cohort. The quantum algorithm can be made recursive or iterative, but is written iteratively for simplicity. The \emph{performMaxcut} subroutine is implemented on a D-Wave Systems hybrid solver \cite{dwavehybrid}. Once a cohort assignment has completed, we can remove edges between students of different cohorts from the network, as we assume that students in separate cohorts can no longer transmit disease through a shared class.

\begin{algorithm}[htb]
\SetAlgoLined
 G = Network Graph\;
 N = Target number of cohorts (power of 2) \;
 currCohorts = [G]\;
 \While{currCohorts.length $<$ N}{
    newCohorts = []\;
    \For{c in currCohorts} {
        c1, c2 = performMaxcut(c)\;
        newCohorts.append(c1)\;
        newCohorts.append(c2)\;
    }
    currCohorts = newCohorts\;
 }
 \caption{Quarantine Partitioning Algorithm: iteratively  perform max-cut until we reach the desired number of cohorts}
 \label{alg1}
\end{algorithm}

\begin{algorithm}[htb]
\SetAlgoLined
 G = Network Graph\;
 N = Target number of cohorts (power of 2) \;
 currCohorts = [N arrays]\;
    \For{i in G.nodes} {
        cohort\_idx = random.choice([1,2,...N])\;
        currCohorts[cohort\_idx].append(i)\;
    }
    \caption{Random Partitioning Algorithm: assign each individual to a cohort with uniform probability}
    \label{alg2}
    
\end{algorithm}

\subsection{Epidemic Modelling}\label{epidemics}
Once a network has been set, a disease outbreak can be simulated using the Susceptible-Infected-Recovered (SIR) model,  derived from the Kermack-McKendrick theory of disease spread \cite{kermack27}. We implemented this model with the Python library NDlib \cite{ndlib}. In order to fully specify the disease model, we needed to define the following parameters:
\begin{itemize}
    \item $r_r$ (recovery rate): the probability that an infected individual will recover in any given day.
    \item $r_i$ (infection rate): the probability that a healthy individual will be infected in any given day.
\end{itemize}
Estimates vary, but show for COVID-19, the average recovery time $T_r$ is around 10 days \cite{recovery}. Since $r_r$ denotes a constant probability of recovering in a single day, the probability of an individual to recover after multiple days follows a geometric distribution, with an average recovery time that should equal $T_r$. Thus given $T_r$, we can calculate $r_r = 1/T_r$.

To calculate $r_i$, we note that in epidemiology, $R_0$ quantifies the average number of people that an individual can infect during the course of an illness. The exact $R_0$ value for COVID-19 depends on a wide array of variables, but has been reported to range from 1.5 to 6.7 \cite{achaiah20}. Using a chosen $R_0$ value, we can estimate $r_i$ via the equation below: 
\begin{equation} \label{eu_eqn}
R_0 = avg\_deg(G)*r_i/r_r
\end{equation}
The reasoning for this is as follows: each infected individual will interact with $avg\_deg(G)$ individuals on average per day, and each interaction may create a new infection with probability $r_i$. Thus, at each time step, the expected number of infections caused by an infected individual in a day is avg\_deg(G)*$r_i$. Moreover, since the infected individual is expected to be infected for 1/$r_r$ days, the total expected number of infections needs to be further multiplied by 1/$r_r$.

Having specified $r_i$, $r_r$, and a CEN or SIN, we can run the SIR disease model and calculate two valuable data points, the \emph{total infected percentage}, representing the percent of the population that contracts the disease at some point by the end of the simulation, and the \emph{peak infected percentage}, which is the percent of infected students at the peak of the outbreak.

\section{Results}\label{results}

\subsection{Course Enrollment Networks}\label{cen}

\begin{figure}[htb]
\centerline{\includegraphics[width=8.5cm]{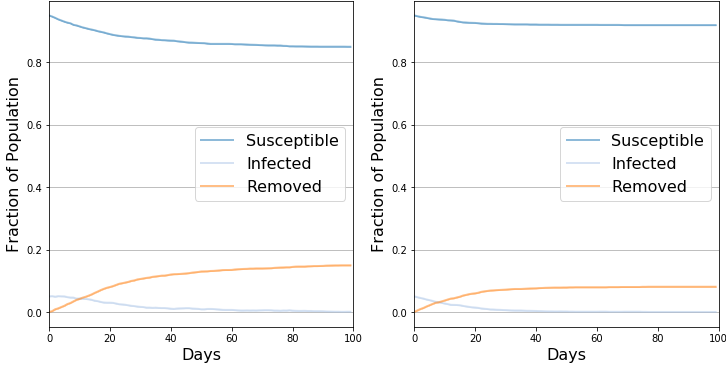}}
\caption{Disease progression comparison between random (left) and QA-assisted (right) student groupings for 3800-student CEN based on Cornell University Ithica Campus. The susceptible, infected, and removed curves represent percentages of students that have not yet gotten the disease, currently have the disease, and have already had the disease, respectively}
\label{cornell_sir}
\end{figure}

For the 3800-student CEN, modeled after Cornell Ithaca campus’s coursework enrollment network, our algorithm was able to reduce the total infected percentage by over 50\% compared to random quarantine assignment with four groups. Moreover, the infection curve steadily declines as opposed to the random quarantine assignment, which produces an infection curve that remains constant for a few weeks before descending. This disease progression is shown in Fig \ref{cornell_sir}. For this simulation, we assume 5\% of students are initially infected, an $R_0$ value of 6, and a recovery time of 10 days.  In addition to this simulation, we tested other initial infected student percentages and number of cohorts. In these simulations, the total infections were reduced between 10-50\%, depending on the two variables mentioned.

\subsection{Student Interaction Networks}\label{sin}

\begin{figure}[htb]
\centerline{\includegraphics[width=9.5cm]{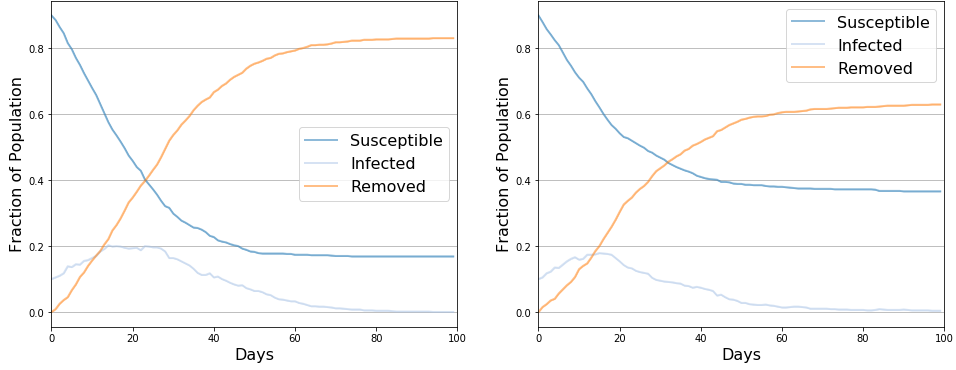}}
\caption{Disease progression comparison between random (left) and QA-assisted (right) student groupings for 800-student SIN based on Harvey-Mudd College Campus. The susceptible, infected, and removed curves represent percentages of students that have not yet gotten the disease, currently have the disease, and have already had the disease, respectively}
\label{harvey_sir}
\end{figure}

For the 800-student SIN, with students grouped into dormitories according to Harvey-Mudd campus housing, our algorithm was able to reduce the total infected percentage by over 20\% compared to random quarantine assignment. This disease progression is shown in Fig \ref{harvey_sir}. In this simulation we assume 1\% of students are initially infected, an $R_0$ value of 2, a recovery time of 10 days, and added floor, dorm, and campus interaction edges with probability .2, .005, and .0001 respectively.

In order to more thoroughly test a wider variety of parameters (initial infection rate, infection rate, and interaction edge rates), we repeated the comparisons between quantum and random groupings with each parameter ranging between lower and upper bounds given by Table \ref{tab2}. Using parameters from these ranges, we found that on average we were able to reduce the peak and total infected percentages by only 2.3\% and 1.7\%, respectively, over random assignments. Furthermore, the quantum groupings provided better infection percentages in only 53\% of the settings. Upon closer analysis, we realized that in the majority of these parameter settings, there was uncontrolled spread in both the random and quantum cohort case, which accounted for the minuscule improvement. We thus hypothesized that quantum groupings would be more effective in parameter regimes corresponding to values of the listed parameters. Motivated by this idea, we set our parameters according to Table \ref{tab3}, ran multiple simulations with randomly added edges, and found that in this case, we were able to reduce the peak and total infected percentages by 4.9\% and 16.7\% compared to random groupings. Additionally, quantum groupings performed better than random groupings in 9 out of the 10 simulations.

\begin{table}[htb]
\caption{Parameter Variation for SINs}
\begin{center}
\begin{tabular}{|c|c|c|}
\hline
\textbf{}&\textbf{\textit{lower bound}}& \textbf{\textit{upper bound}} \\
\hline
\textbf{initial infection}&\textbf{\textit{0.05}}& \textbf{\textit{0.5}} \\
\hline
\textbf{infection rate}&\textbf{\textit{0.01}}& \textbf{\textit{0.1}} \\
\hline
\textbf{floor interaction edge rate}&\textbf{\textit{0.1}}& \textbf{\textit{0.6}} \\
\hline
\textbf{dorm interaction edge rate}&\textbf{\textit{0.001}}& \textbf{\textit{0.016}} \\
\hline
\textbf{campus interaction edge  rate}&\textbf{\textit{0.00005}}& \textbf{\textit{0.0005}} \\
\hline
\end{tabular}
\label{tab2}
\end{center}
\end{table}

\begin{table}[htb]
\caption{Parameters for low-density SINs}
\begin{center}
\begin{tabular}{|c|c|}
\hline
\textbf{}&\textbf{\textit{value}}\\
\hline
\textbf{initial infection}&\textbf{\textit{0.05}} \\
\hline
\textbf{infection rate}&\textbf{\textit{0.02}} \\
\hline
\textbf{floor interaction edge rate}&\textbf{\textit{0.2}} \\
\hline
\textbf{dorm interaction edge rate}&\textbf{\textit{0.005}} \\
\hline
\textbf{campus interaction edge rate}&\textbf{\textit{0.0001}} \\
\hline
\end{tabular}
\label{tab3}
\end{center}
\end{table}

\section{Discussion}\label{discussion}
The aim of this work was to introduce a technique that leverages the computational power of quantum annealing to group students on college campuses into physically separated cohorts in order to reduce the spread of COVID-19. This was achieved via the recursive application of Quantum Annealing (QA) to course enrollment networks (CENs) and more general student interaction networks (SINs), both of which capture physical interactions between students. We found that for both CENs and SINs, with certain assumptions on disease spread, the proposed method was able to reduce the total and peak infected percentages under the SIR epidemic model compared to random cohort assignment. A severe limitation to our method, however, is that our method is not effective if the combination of initial infected student percentage, infection rate, and added interaction edges in a SIN is large enough that there is uncontrolled spread in both the quantum and random grouping case. We also note that we had to estimate values of floor, dorm, and campus interaction edges for the SIR model, as we were not able to find publicly available estimates of these values. Additionally, there may be more interactions in a SIN that cannot be accounted for, such as interactions with individuals living off-campus.

As mentioned, the advantages and possible speedups of QA over classical Max-Cut algorithms and heuristics remain widely unknown for arbitrary graphs. Thus, one could immediately build off this work by viewing the recursive Max-Cut step as a black box, and deciding groupings using purely classical, quantum gate-based, or hybrid quantum-classical devices for optimization. This work could then be relevant outside of quantum annealing literature. One could additionally replace recursive Max-Cut with Max k-Cut, wherein one divides the problem graph into $k$ cohorts in a single optimization procedure. In order to demonstrate quantum advantage, one must be able to regularly show that the groupings returned from the quantum-assisted method outperform groupings returned by the best possible classical algorithms or heuristics for an array of CENs or SINs. This would most likely require Max-Cut solved on a quantum annealer to outperform any classical method, and current research into this comparison is inconclusive. As stated in the introduction, our goal was not to perform this extensive benchmarking, but rather introduce a novel application of QA.

This approach can likewise be extended to any environment that can be described by a mostly self-contained network representing individuals and shared physical environments, such as office spaces, gyms, and primary/secondary schools.

\section*{Acknowledgment}

J.S. and V.L. would like to thank enlightening and helpful discussions from Dr. Hossein Sadeghi, as well as D-Wave Systems, INC, for free COVID-19 research access to D-Wave Leap, providing cloud-accessed quantum annealers and quantum/classical hybrid solvers. J.S. and V.L. would also like to thank the Quantum Computing @ Berkeley Club, and Dr. Bjoern Hartmann, our mentor for the 2020 Jacobs Institute Of Design Innovation Covid-19 Design Challenge\cite{jacobs}. J.S. would also like to thank Maya Chan Morales for research consultation and editing. All code used in this project can be found at https://github.com/qcberkeley/optimization

\end{document}